\documentclass[twoside,a4paper,11pt]{sca}
\usepackage{graphicx}
\usepackage{hyperref}
\usepackage{natbib}  
\begin{document}
\pagenumbering{arabic}
\pagestyle{myheadings}
\thispagestyle{empty}
{\flushright\includegraphics[width=\textwidth,bb=90 650 520 700]{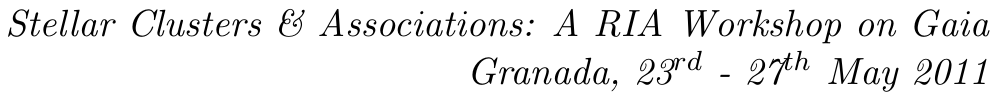}}
\vspace*{0.2cm}
\begin{flushleft}
{\bf {\LARGE
%
The Milky Way Nuclear Star Cluster 
%
}\\
\vspace*{1cm}
%
R. Capuzzo-Dolcetta$^{1}$,
F. Antonini$^{2}$, 
and 
A. Mastrobuono-Battisti$^{1}$
%
}\\
\vspace*{0.5cm}
%
$^{1}$
Dep. of Physics, Sapienza, Univ. di Roma, Italy\\
$^{2}$
Dep. of Physics, Rochester Institute of Technology, Rochester, USA\\
%
\end{flushleft}
%
\markboth{
The Milky Way Nuclear Cluster
}{ 
%
R. Capuzzo-Dolcetta et al. 
%
}
\thispagestyle{empty}
\vspace*{0.4cm}
\begin{minipage}[l]{0.09\textwidth}
\ 
\end{minipage}
\begin{minipage}[r]{0.9\textwidth}
\vspace{1cm}
\section*{Abstract}{\small
%
In the center of the Milky Way, as well as in many other galaxies, a compact star cluster around a very massive black hole is observed. One of the possible explanations for the formation of such Nuclear Star Clusters is based on the `merging' of globular clusters in the inner galactic potential well. By mean of sophisticated N-body simulations, we checked the validity of this hypothesis and found that it may actually has been the one leading to the formation of the Milky Way Nuclear Star Cluster.

%
\normalsize}
\end{minipage}
%
%
%
\section{Introduction}
The existence of very compact star clusters around the centers of many galaxies is well ascertained, nowadays. These clusters, called  Nuclear Star Clusters (NSCs) are among the densest star clusters observed, with effective radii of a few pc and central luminosities up to $10^7$ L$_\odot$. Recently, the study of dense stellar systems has raised new interest because of the discovery, in an ever increasing number of galaxies, of compact nuclei in form of stellar resolved systems and/or NSCs which are now known to be present in galaxies across the whole Hubble sequence and not only in the dE, N galaxies \citep{bek10}. A radial profile and velocity structure for a NSC can be determined only for the Milky Way NSC which is close enough  that it can be resolved into individual stars. The Milky Way NSC has an estimated mass of $10^7$ M$_\odot$, and it hosts a massive black hole whose mass, $4.3\times 10^6$ M$_\odot$, is uniquely well determined. 
\\ The formation mechanism of nuclear star clusters  is unknown. Two competing models are possible. In the gas model, a NSC can form from the gas that migrates to the center of the galaxy where then forms stars. A variety of scenarios have been proposed to account for the required fast radial inflow of gas into the galactic center, including the magneto-rotational instability in a differentially rotating gas disk, tidal compression in shallow density profiles or dynamical instabilities. Alternatively, in the merger model, massive clusters migrate to the center via dynamical friction and merge to form a dense nucleus \citep{tre75,capdol93}. Observations of NSCs in dE galaxies suggest that the majority, but not all, dE nuclei, could be the result of packing mass in form of orbitally decayed globular clusters (GCs) \citep{lot04}. Numerical simulations have also shown that the basic properties of NSCs, including their shape, mass density profile, and mass-radius relation are reproduced in the merger model under a variety of explored conditions \citep{capdol08a,capdol08b,Har11}.

\section{The Milky Way nuclear cluster}
As we said, a NSC is present in the central region of the Milky Way. For obvious reasons, it is the only NSC which can be resolved into individual stars in spite of its high density \citep{sch07}; moreover, detailed  kinematic studies exist that allows a precise determination of the mass of the central black hole, M$_{BH} =4.3\times 10^{6}$ M$_\odot$. Moreover, the relaxation time at Sgr A$^*$ influence radius is robustly estimated as $20$ -- $30$ Gyr \citep{mer10} suggesting there has not been time for a Bahcall-Wolf cusp to rise. This is consistent with the almost flat density distribution of late type stars within 0.5 pc from the center \citep{buc09}. As stated in the Introduction, at least two competing models are possible for the formation of the MW NSC. These mechanisms must account for the observed  evidence of the MW NSC, i.e.: a mass of the NSC M$_{NSC}\sim 10^7$ M$_\odot$, a density profile with a {\it core} of about 0.5 pc and decreasing as $r^{-1.8}$ up to $r\simeq 30$ pc.
Out of 30 pc there is a large nuclear stellar and molecular disk of approximately same radius.

\subsection{Globular cluster merging and MW NSC formation}
We modeled the final evolution of a set of massive GCs which decayed in the inner region of the Milky Way due to the dynamical friction braking exerted by the galactic stellar field. The details of the computations and the extended presentation and discussion of results are given in \cite{ant11}; here we give just a brief summary. 
\\  Our Galaxy, likely triaxial in its inner part, is dense enough to make the orbits of GCs with mass around $10^6$ M$_\odot$ to  decay (by dynamical friction) to the inner 50 pc in a time significantly shorter than a Hubble time. We thus consider 12 such GCs and their final evolution toward a merger state studied via N-body simulations of their motion in the inner region of the Milky Way, represented self-consistently as a set of N particles. The N body galactic environment is sampled from a shallow ($\rho \propto r^{-1/2}$) cusp; the contribution of the central massive black hole is also considered. Simulations are done with both the high precision, parallel PhiGRape \citep{har07} and NBSymple \citep{cap11} codes running on the RIT Grape cluster and on the CPU+NVIDIA TESLA C2050 platform in Roma, Sapienza.

\section{Results}
\begin{figure}
\center
\includegraphics[width=8.3cm,angle=0,clip=true]{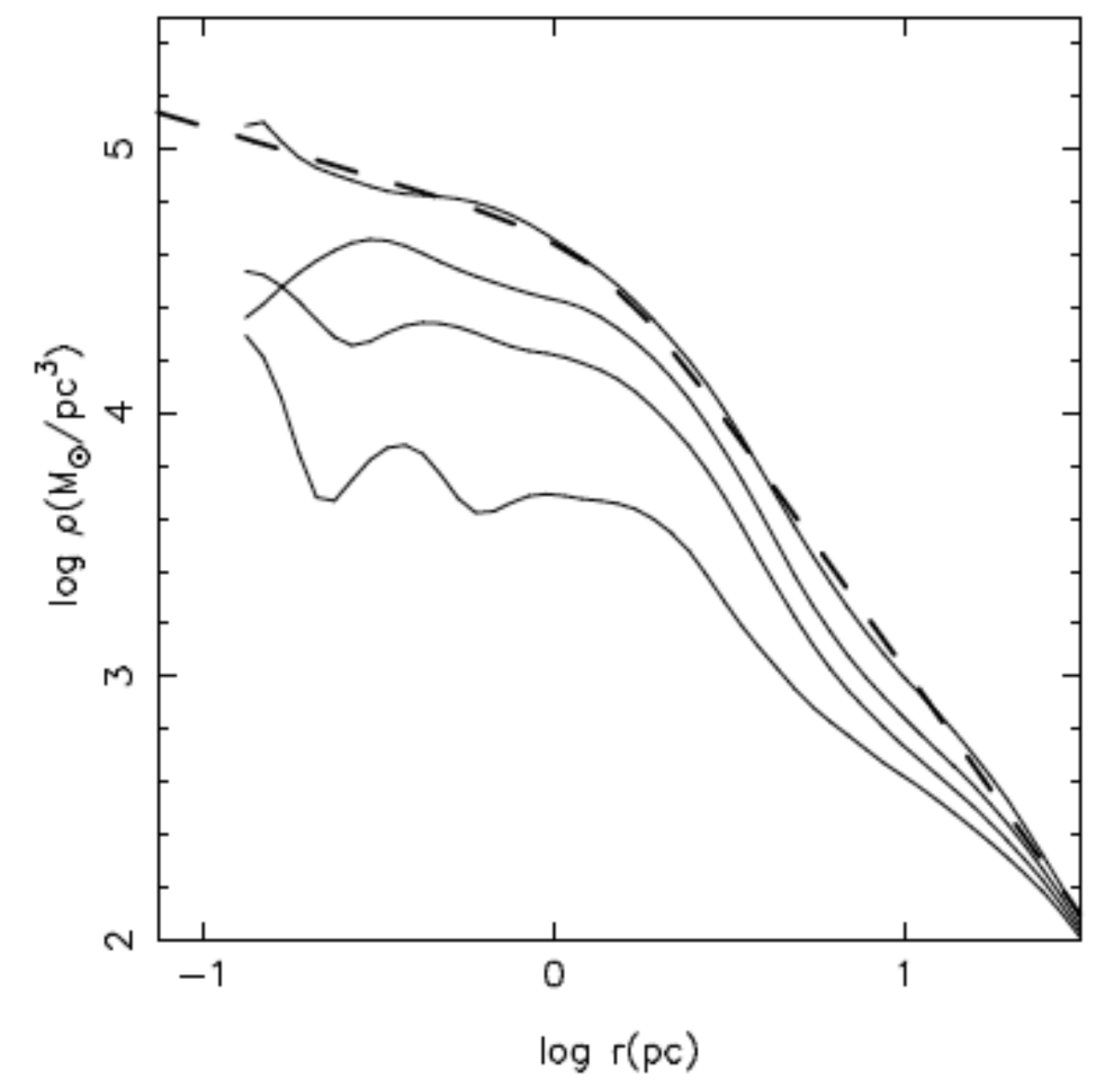} ~
\caption{\label{fig1} Spatial profile of the central NSC after 3, 6, 9 and 12 mergers. The central density grows with time. The dashed line
is the fit to the NSC profile obtained at the end of the entire simulation using a broken power law model.
}
\end{figure}

\begin{figure}
\center
\includegraphics[scale=0.35]{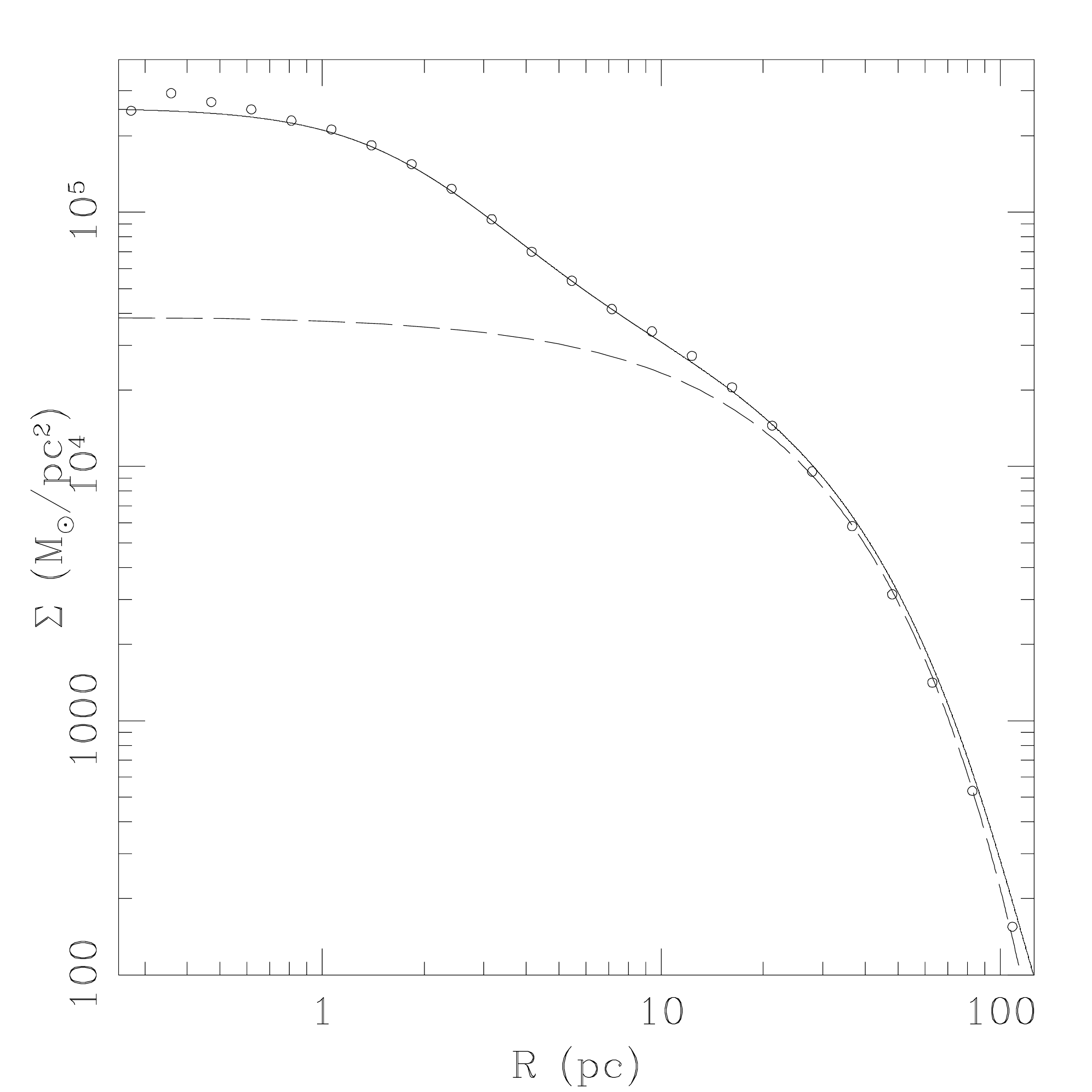} ~
\includegraphics[scale=0.35]{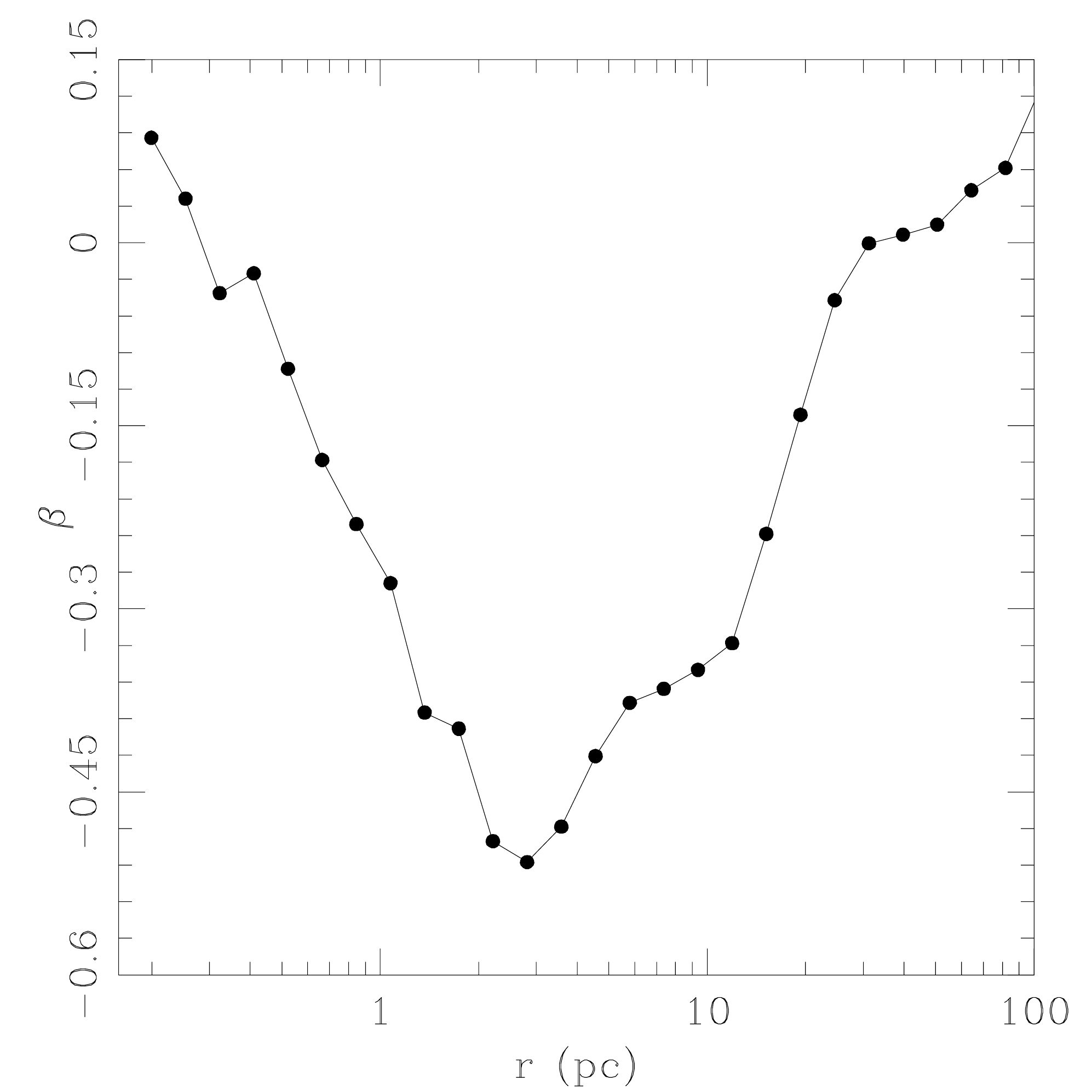}
\caption{\label{fig2lr} Left panel: projected density profile of the merged system at the end of the simulation (solid line). Dashed line refers to the galactic density profile. Right panel: radial behaviour of the anysotropy parameter $\beta=1-\sigma_t^2/\sigma_r^2$.}
\end{figure}

The merging occurs rather quickly: after about 20 crossing times the resulting system attains a quasi equilibrium configuration, as it is shown by the almost steady time profile of Lagrangian radii. This corresponds to a slowly evolving (via relaxation) super stellar cluster (SSC) which oscillates around the massive black hole. 
Figure \ref{fig1} shows the growth with time of the spatial density profile after the various merger events. The core structure is conserved. Figure \ref{fig2lr} (left panel) gives the comparison between the surface density of the 12 GCs merger and the underlying galactic stellar profile. 
The kinematic properties of the merger system are resumed by the Fig. \ref{fig2lr} (right panel), which shows the presence of a tangentially biased zone, intermediate between an inner (close to the massive BH, $r\leq 0.3$ pc) and outer ($r\geq 40$ pc) isotropic regions. This tangential anisotropy grows with the number of infalls. This may be an important signature of the modes of NSC formation.

The SSC has a density which is not far from the sum of the individual GC densities, so that the projected density profile in the nuclear region is remarkably similar to that of the Milky Way NSC. 
The final NSC morphology is nearly oblate with a value $T\sim 0.2$ for the triaxiality parameter. 

\section{Conclusions}
As already shown by previous papers \citep{capdol93,capdol08a,capdol08b} the dynamical friction dragging of massive globular clusters toward the galactic center is a viable explanation of the local high densities. The novelty of this work is the inclusion in the simulations of a massive black hole in the galactic center. Its role, although important in the vicinity of its influence radius, does not alter the general characteristics of the merger event of a set of GCs falling to the center. The merger reaches a sort of quasi-steady state, slowly evolving due to internal relaxation which is partly affected by the presence of the massive black hole.
With regard to the Milky Way nuclear cluster, it is relevant noting how the original `core' profile of the GCs (the NSC building blocks) is maintained for a time long enough to justify the core actually observed in the MW NSC. Moreover, the outward profile radial slope, $\rho \propto r^{-2}$, of the merger system is in good agreement (within 30 pc from the center) with the observed  MW NSC profile. Surely, an even more stringent answer to the question of the MW NC formation will come from a high resolution, well extended in time numerical study of the secular evolution of the merger remnant as well as from a detailed study of the expected observed stellar population (both in progress).
%
%
\small  
%
%

%
%
%
%
%


\begin{thebibliography}{}

\bibitem[Antonini et al.(2011)]{ant11}Antonini, F., Capuzzo-Dolcetta, R., Mastrobuono-Battisti, A., \& Merritt, D. 2011, in preparation


\bibitem [Bekki \& Graham(2010)]{bek10} Bekki, K. \& Graham, A.~W.  2010, ApJ, 714, 313


\bibitem[Buchholz et al.(2009)]{buc09}Buchholz, R. M., Sch\"{o}del, R., \& Eckart, A. 2009, A\&A, 499, 483

\bibitem[Capuzzo-Dolcetta(1993)]{capdol93} Capuzzo-Dolcetta, R., 1993, ApJ, 415, 616

\bibitem[Capuzzo-Dolcetta et al.(2008a)]{capdol08a} Capuzzo-Dolcetta, R., Miocchi, P., 2008a, MNRAS, 388, L69
\bibitem[Capuzzo-Dolcetta et al.(2008b)]{capdol08b} Capuzzo-Dolcetta, R., Miocchi, P., 2008b, ApJ, 681, 1136

\bibitem[Capuzzo-Dolcetta et al.(2011)]{cap11} Capuzzo-Dolcetta, R., Mastrobuono-Battisti, A., Maschietti, D., 2011, NewAstron, 16, 284




\bibitem[Harfst et al.(2007)]{har07} Harfst, S., Gualandris, A., Merritt, D., Spurzem, R., Portegies Zwart, S., \& Berczik, P. 2007, NewAstron, 12, 357

\bibitem[Hartmann et al.(2011)]{Har11} Hartmann, H., Debattista,  V.~P., Seth, A., Cappellari, M., \& Quinn, T.~R., \ 2011, 

\bibitem[Lotz et al.(2004)]{lot04}Lotz J. M., Miller B. W., Ferguson H. C., 2004, ApJ, 613, 262

\bibitem[Merritt(2010)]{mer10} Merritt, D., 2010, ApJ, 718, 739


\bibitem[Sch\"odel et al.(2007)]{sch07} Sch\"odel, R., Eckart, A., Alexander, T., et al. 2007, A\&A, 469, 125

\bibitem[Tremaine et al.(1975)]{tre75} Tremaine, S. D., Ostriker, J. P., \& Spitzer, L., Jr. 1975, ApJ, 196, 407


\end{thebibliography}

\clearpage

\end{document}